\documentclass[pdftex,twocolumn,epjc3]{svjour3}
\RequirePackage[T1]{fontenc}
\smartqed
\RequirePackage{graphicx}
\RequirePackage{mathptmx}
\RequirePackage{flushend}
\RequirePackage[numbers,sort&compress]{natbib}
\RequirePackage[colorlinks,citecolor=blue,urlcolor=blue,linkcolor=blue]{hyperref}

\usepackage{newtxtext,newtxmath}
\DeclareMathOperator*{\argmin}{\arg\min}
\DeclareMathOperator*{\argmax}{\arg\max}
\usepackage{xspace}
\newcommand{\btheta}{\ensuremath{\boldsymbol{\theta}}\xspace}
\newcommand{\bx}{\ensuremath{\mathbf{x}}\xspace}

\journalname{Eur. Phys. J. C}

\begin{document}

\title{Mixture Density Network Estimation of Continuous Variable Maximum Likelihood Using Discrete Training Samples}

\author{Charles Burton\thanksref{e1,addr1}
        \and
        Spencer Stubbs\thanksref{e2,addr1,note1}
        \and
        Peter Onyisi\thanksref{e3,addr1}
}

\thankstext{e1}{e-mail: burton@utexas.edu}
\thankstext{e2}{e-mail: f.spencer.stubbs@gmail.com}
\thankstext{e3}{e-mail: ponyisi@utexas.edu}
\thankstext{note1}{Now at Physics Department, Rutgers University.}

\institute{
    Department of Physics, University of Texas, Austin, TX, USA\label{addr1}
}

\date{Received: date / Accepted: date}

\maketitle

\begin{abstract}
Mixture Density Networks (MDNs) can be used to generate posterior density functions of model parameters $\boldsymbol{\theta}$ given a set of observables $\mathbf{x}$. In some applications, training data are available only for discrete values of a continuous parameter $\boldsymbol{\theta}$. In such situations, a number of performance-limiting issues arise which can result in biased estimates. We demonstrate the usage of MDNs for parameter estimation, discuss the origins of the biases, and propose a corrective method for each issue.
\end{abstract}

\section{Introduction}
A frequent goal in the analysis of particle physics data is the estimation of an underlying physical parameter from observed data, in situations where the parameter is not itself observable, but its value alters the distribution of observables. One typical approach is to use maximum likelihood estimation (MLE) to extract values of the underlying parameters and their statistical uncertainties from experimental distributions in the observed data. In order to do this, a statistical model $p \left( \bx | \btheta \right)$ of the observable(s), as a function of the underlying parameters, must be available. These are frequently available only from Monte Carlo simulation, not from analytic predictions. In typical usage, the value of a parameter is varied across a range of possible values; the derived models (determined from histograms or other kernel density estimators, or approximated with analytic functions) are then compared to the distributions in the observed data to estimate the parameter.

A number of methods to perform this type of inference have been discussed in the literature. See, for example, Refs.~\cite{Brehmer:2018eca,Flesher:2020kuy,Andreassen:2020gtw,Baak_2015,Cranmer:histfactory,Read:histinterp}. Some of these also use machine-learning approaches, and many support the use of multiple observables in order to improve statistical power.

If one has a complete statistical model $p \left( \bx | \btheta \right)$ for the observables available for any given value of the parameter, the MLE can be computed. Unfortunately, this is usually difficult to determine analytically, especially if there are multiple observables with correlations, detector effects, or other complications. An alternative procedure is to directly approximate the likelihood function of the parameter, $\mathcal{L} \left( \btheta | \bx \right)$.

Mixture Density Networks~\cite{Bishop94} solve a closely-related task. They are used to approximate a posterior density function $p \left( \btheta | \bx \right)$ of parameters \btheta from input features \bx as a sum of basis probability density functions (PDFs). More specifically, the neural network predicts the coefficients/parameters of the posterior density. With Bayes' theorem, we will relate the posterior density, which is output by the network, to the desired parameter likelihood function. Notably, because of the flexibility of network structure, MDNs permit the straightforward use of multidimensional observables, as well as approximating the posterior density function with any desired set of basis PDFs.

When training MDNs, one typically assumes that all input parameter values are equally likely to be sampled in the training dataset, and that the parameter is continuous. In essence, this is equivalent to specifying a flat prior for the application of Bayes' theorem that translates the posterior density that the MDN learns into the likelihood function.

However, such datasets may not be readily available for various reasons: one may share Monte Carlo samples with analyses using other estimation techniques which use histograms at specific parameter values to build up templates, or there may be computational difficulties with changing the parameter values in the Monte Carlo generator for every event. For example, in a top quark mass measurement, Monte Carlo event generators do not efficiently support the case of simulating events along a continuum of possible top mass values. Rather, events are generated where the top mass has been set to one value on a grid of possible parameter values. In this work, we discuss issues which arise when using training samples with discrete parameter values, rather than a continuous parameter distribution, to estimate parameters with MDNs.

A related issue arises when a restricted range of parameter values is used in the training, in which the trained network is reluctant to predict values near the boundaries of the range due to the relative lack of training events in that vicinity. This occurs even when training with a continuous parameter distribution and affects tasks other than likelihood estimation, such as simple regression. Since this issue will appear in any practical application of an MDN to estimate a physical parameter, we will discuss it.

The aim of the paper is to demonstrate the construction of MDNs for estimating continuous parameters from datasets populated only with discrete parameter values, and to discuss pitfalls that can occur in the training. This paper is structured as follows. The basic concept of Mixture Density Networks is introduced and the application to likelihood estimation is discussed. Potential biases in the network training are explained, and procedures to mitigate them are proposed, in the context of specific examples. The performance of trained MDNs is demonstrated. Finally, limitations of the technique and avenues for future improvement are discussed.

\begin{figure}
    \centering
    \includegraphics[width=0.4\textwidth]{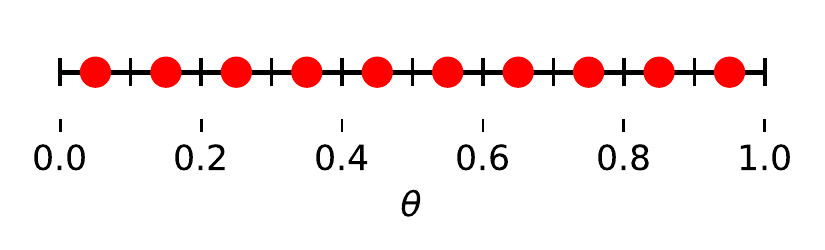}
    \caption{For some applications, training data are only available at discrete values of $\theta$ (e.g. at each of the red markers). For every example discussed here, the training samples consist of 10 template data sets with parameter $\theta$ equally spaced between 0 and 1.}
    \label{fig:template_thetas}
\end{figure}

\section{Mixture Density Networks as Likelihood Approximators\label{sec:mdn}}
A Mixture Density Network is a neural network where the output is a set of parameters of a \textit{function}, defined to be a properly normalized PDF. The outputs of the MDN can be, for example, the mean, width, and relative contributions of a (fixed) number of Gaussian distributions. But, generally speaking, the goal of MDN usage is to obtain an estimate of a posterior density function $p \left( \btheta | \bx \right)$ of a data set that includes parameters \btheta and observations \bx.

The MDN represents the target posterior density with a weighted sum of $n$ generic basis PDF functions $\mathcal{B}_{n}$,
\begin{align} \label{eq:likelihood_basis}
    \tilde p \left( \btheta | \bx ; \mathbf{Z} \right) = \sum_{i=1}^{n}{c_i \left( \bx \right) \cdot \mathcal{B}_{i} \left( \boldsymbol{\theta}; \mathbf{z}_{i} \left( \bx \right) \right)},
\end{align}
where $c_i \left( \mathbf{x} \right)$ and $\mathbf{z}_i \left(\mathbf{x} \right)$ are the $\mathbf{x}$-dependent coefficients and parameters of the basis functions, which are predicted by the network, and $\mathbf{Z} = \left\{ c_i \right\} \cup \left\{ \mathbf{z}_i \right\}$ is shorthand to represent all of the coefficients of the learned model. Typically, the conditions $c_i \in [0,1]$ and $\sum_i c_i = 1$ are imposed, for example through the softmax function. In principle, the basis functions can be any set of basis PDFs. A useful choice for many applications is a mixture of $n$ Gaussian functions. (Since we are estimating a posterior density function of a continuous parameter, we might expect a minimum in the negative logarithm of this function. The minimum would be quadratic to leading order, making Gaussians a natural choice for PDF basis functions.) For that choice, the neural network's output is a set of $3n-1$ independent coefficients $\mathbf{Z} \left( \bx \right) = \left\{ c_{i}\left(\mathbf{x}\right), \mu_{i}\left(\mathbf{x}\right), \sigma_{i}\left(\mathbf{x}\right) \right\}$ (with one softmax normalization condition).

To train the network, in each epoch, the output function of the network $\tilde p \left( \btheta | \bx ; \mathbf{Z} \right)$ is evaluated for each point in the training data set $ \left\{ \left( \boldsymbol{\theta}_{j}, \mathbf{x}_{j} \right) \right\}$. The cost function,
\begin{align}\label{eq:likelihood}
    \mathcal{C} \left( \mathbf{Z} \right) = - \log \left[ \prod_{j=0}^{m} \tilde p \left( \btheta_{j} | \bx_{j} ; \mathbf{Z} \left( \bx_j \right) \right) \right],
\end{align}
is the negative logarithm of the product of these values. The error is backpropogated through the network in the standard way, and the cost is minimized to determine the ideal coefficients,
\begin{align}
    \mathbf{\hat Z} = \argmax_{\mathbf{Z}} \left\{ \mathcal{C} \left( \mathbf{Z} \right) \right\}.
\end{align}

From the physics standpoint of parameter estimation with discretized data, we are not actually interested in using the network to model the \emph{true} posterior density, as one might typically do in MDN applications. (With discrete training samples, the true posterior of the training data set involves delta functions at the various values of \btheta where each template lies.) Instead, we convert $\tilde p \left( \btheta | \bx ; \hat{\mathbf{Z}} \right)$ created by the MDN into an estimate of the likelihood function $\mathcal{L} ( \btheta | \bx )$. Using Bayes' theorem,
\begin{align}\label{eq:bayes}
    \mathcal{L}(\btheta|\bx) = \tilde p \left( \bx | \btheta ; \hat{\mathbf{Z}} \right) = \frac{\tilde p \left( \btheta | \bx ; \hat{\mathbf{Z}} \right) p \left( \bx \right)}{p \left( \btheta \right)}.
\end{align}
Notably, in the mindset of estimating some parameters \btheta, as long we ensure a flat prior $p \left( \btheta \right)$, the likelihood that we seek and the posterior density which is output by the trained MDN differ only by an irrelevant multiplicative factor---the prior $p \left( \bx \right)$, which is determined by the entire training set, and does not depend on \btheta.

In order for the MDN to effectively \emph{interpolate}, it is important that there not be too much freedom in the MDN output. For example, suppose there were an equal number of training templates and Gaussian components in $\tilde p \left( \btheta | \bx \right)$. The network could then essentially collapse the Gaussian functions to delta functions at each template \btheta value and reduce the cost function without limit. As long as the observed values \bx can reasonably be produced by multiple values of \btheta, and the number of basis functions in the MDN is kept reasonably low, the MDN will naturally be forced to interpolate between parameter points, as desired for estimating $\mathcal{L} \left( \btheta | \bx \right)$.

\section{Density of Parameter Points}\label{s:density_of_points}
The application of Bayes' theorem in Eq.~\eqref{eq:bayes} involves $p(\btheta)$, and simplifies if $p(\btheta)$ can be assumed to be flat. To ensure this condition, the locations of the templates in parameter space should be chosen to ensure an equal density of training points across the entire range.

This first requires that the templates must be equally distributed in the parameter space. Otherwise, the density of training points would be non-constant, implying a non-flat $p \left( \btheta \right)$. This would bias the network.

Secondly, this necessitates that the parameter range extend slightly outside of the range of the templates in parameter space. For example, suppose we have one parameter $\theta$ and the parameter range is selected as $\theta \in \left[ 0, 1 \right]$, and 10 templates are to be used. To ensure equal and unbiased coverage of the parameter range, they should be placed at $\theta = 0.05, 0.15, \ldots, 0.95$, as shown in Fig.~\ref{fig:template_thetas}. If the extra gaps (from 0 to 0.05 and from 0.95 to 1) are not included at the edges (for example, if eleven templates were used, at $\theta = 0, 0.1, \ldots, 1$), then the density of training data is higher at the extremal values of $\theta$, again creating a non-flat $p \left( \theta \right)$ and a bias in the training.

Essentially, the templates' parameter values should lie at the centers of equal-width histogram bins that extend from the lowest to highest values of $\theta$. Note that, in Fig.~\ref{fig:template_thetas}, each $\theta$ bin has an equal amount of training data. This condition can be generalized to multidimensional parameter spaces.

\section{PDF Normalization}\label{s:pdf_normalization}
Another issue arises with respect to the normalization of the MDN output. The MDN's output posterior $\tilde p \left( \btheta | \bx ; \mathbf{Z} \right)$ might not be constructed with any knowledge of the range of parameters \btheta in the training set. For example, Gaussian basis functions have infinite support and therefore will always predict a non-zero posterior density outside the range of the training data. Proper training of an MDN requires that the output posterior density be properly normalized across the selected range of $\boldsymbol{\theta}$ for MDN training to work properly. If this is not done, parameter values near the edges of the range will be penalized because the posterior predicts parameter values to occur outside of the range, and these are never encountered in the training data.

For a one-dimensional parameter $\theta \in \left[ \theta_\textrm{min}, \theta_\textrm{max} \right]$, one must require that
\begin{align*} \label{normalization_condition}
    \int_{\theta_\textrm{min}}^{\theta_\textrm{max}} p \left( \theta | \bx \right) d\theta = 1.
\end{align*}
This constraint can be achieved by dividing MDN's predicted posterior density by the integral of the posterior density over the parameter range. Since, during training, the posterior density is never evaluated outside of this range, this results in the proper normalization. 

Practically speaking, it is easier to apply this renormalization procedure for each function $\mathcal{B}_i$ than to do it on the sum. This has the benefit that the condition $\sum_i c_i = 1$ is still valid. In the one-dimensional parameter case, if the cumulative distribution function (CDF) is available,
\begin{align*}
    \mathcal{B}_i \left( \theta; \mathbf{z}_i \right) &= \frac{ \textrm{PDF} \left( \theta; \mathbf{z}_i \right) }{ \textrm{CDF} \left( \theta_\textrm{max}; \mathbf{z}_i \right) - \textrm{CDF} \left( \theta_\textrm{min}; \mathbf{z}_i \right) }.
\end{align*}
This effect is not specific to training with discrete parameter choices, and will generally occur in regions where observed data could be compatible with parameters outside the training range.

\section{Edge Bias from Training on Templates}
We now discuss a bias which arises from the discreteness of the input parameter values. As discussed in the previous sections, in order to achieve a flat $p(\btheta)$, we need to consider the input range of parameters $\btheta$ to be broader than just the range where training data are located. However the cost function is only evaluated on the training data, and so the optimizer can ``cheat'' by overpredicting values of \btheta that are compatible with a broad range of observations, while underpredicting extremal values of \btheta that are not represented in the training data. The symptom of this is that the probability density $\tilde p \left( \bx | \btheta \right)$ implied by the MDN output posterior $\tilde p \left( \btheta | \bx \right)$,
\begin{align*}
    \tilde p \left( \bx | \btheta \right) = \frac{\tilde p \left( \btheta | \bx \right) p \left( \bx \right)}{p \left( \btheta \right)},
\end{align*}
when integrated over data $\bx$, does not integrate to one for all values of \btheta (as one would expect for a properly normalized density function). Rather, it is smaller than one at extremal values of \btheta and greater than one in the interior of the range.

\begin{figure}
    \centering
    \includegraphics[width=0.48\textwidth]{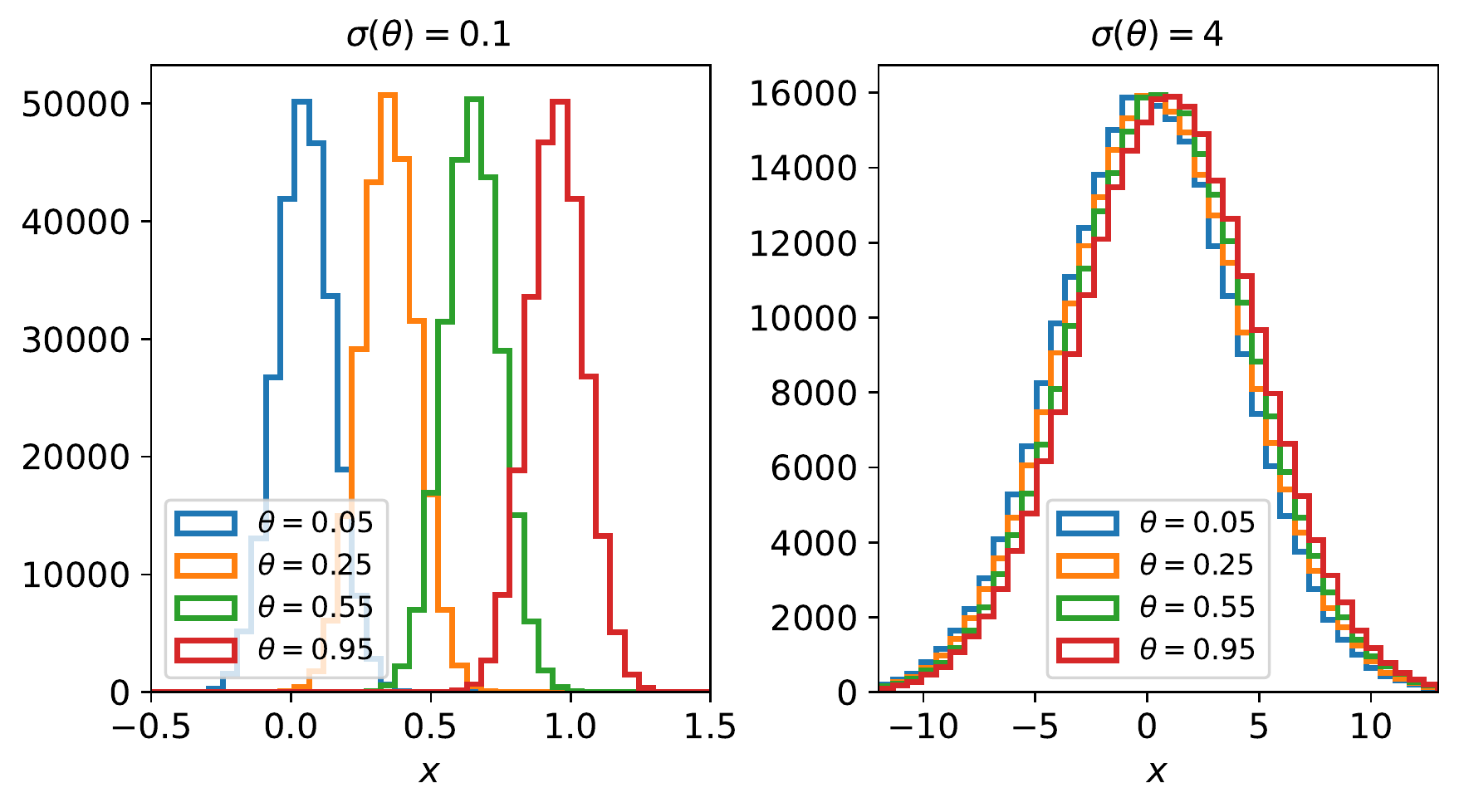}
    \caption{Examples of templates (training data for discrete parameter values) of an observable $x$ for different values of a parameter $\theta$. When the observable templates are distinguishable (left), the edge bias (see text) is negligible. When the templates are harder to distinguish over the range of $\theta$ (right), the correction for the edge bias is critical.}
    \label{fig:fit_templates}
\end{figure}

The size of this effect depends on how much the templates overlap. Fig.~\ref{fig:fit_templates} shows examples of distributions which will demonstrate negligible and extreme bias.

\subsection{Demonstration With Functional Fit}\label{s:biasCorrection}
It should be emphasized that the observed edge bias is \emph{not} unique to the Mixture Density Network method. Rather, it is simply a result of minimizing the cost function Eq.~\eqref{eq:likelihood} composed of the posterior density $p \left( \boldsymbol{\theta} | \mathbf{x} \right)$ for a finite number of templates. To illustrate this, we will show the existence and correction of the bias in a simple functional fit.

Consider a statistical model which, given some parameter value $\theta \in [0, 1]$, produces a Gaussian distribution of $x$, a univariate variable: this could correspond to a ``true'' value $\theta$ and an ``observed'' value $x$ which is subject to resolution effects. For this example, we choose the distribution
\begin{equation} \label{eq:distribution}
    p(x|\theta) = \frac{1}{\sqrt{2 \pi} \sigma} \exp \left( - \frac{1}{2}  \left( \frac{x - \theta}{\sigma} \right)^2 \right),
\end{equation}
with $\sigma = 4$. We choose 10 equally-spaced values of $\theta$ between $0$ and $1$ at which to sample the model to generate training points. Since the width of each template $\sigma$ is much broader than the total range of $\theta$, the 10 templates are not very distinguishable from one another. A few of the templates are shown on the right-hand side of Fig. \ref{fig:fit_templates}.

Next, we attempt to reconstruct the joint probability distribution $p(x, \theta)$ by performing an unbinned maximum likelihood fit of a three-parameter function to the generated data,
\begin{equation*}
    f \left( x, \theta; \mu_m, \mu_b, \sigma \right) = \mathcal{A} \exp\left( -\frac{1}{2} \left( \frac{x - \left(\mu_m \theta + \mu_{b}\right)}{\sigma} \right)^2 \right),
\end{equation*}
where $\mathcal{A}$ is a normalization factor. The best-fit values of the function parameters are not the ones used to generate the data.

\begin{table}
    \centering
    \begin{tabular}{l|r|r|r}
         & $\mu_m$ & $\mu_b$ & $\sigma$ \\\hline
        True values & 1 & 0 & 4 \\
        % $\max \{ \mathcal{L} \}$ & $0.17 \pm 0.03$ & $0.41 \pm 0.02$ & $1.65 \pm 0.12$ \\
        % Unbiased & $1.03 \pm 0.02$ & $-0.02 \pm 0.01$ & $3.99 \pm 0.01$
        $\max \{ \mathcal{L} \}$ & $0.17$ & $0.41$ & $1.65$ \\
        Unbiased & $1.03$ & $-0.02$ & $3.99$
    \end{tabular}
    \caption{The maximum likelihood estimators for the parameters of the function $f$, when determined using discrete parameter inputs, do not match the intended model parameters due to edge effects. Adding a constraint during likelihood optimization resolves this, and the results are consistent with the desired values.}
    \label{tab:fitresults}
\end{table}

\begin{figure}
    \centering
    \includegraphics[width=0.4\textwidth]{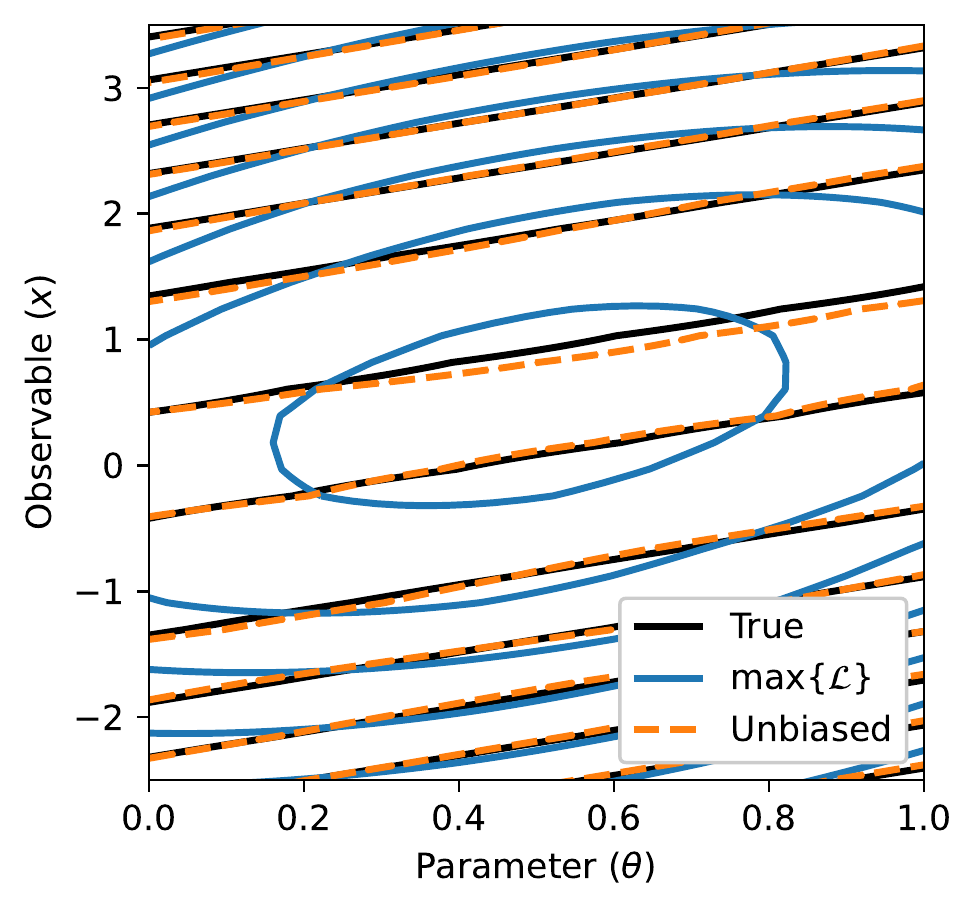}
    \caption{Joint PDFs for the example described in \S\ref{s:biasCorrection}. The ``true'' joint PDF is shown in black. Fitting to the training dataset with discrete parameter values results in the blue PDF, which is different from the desired form. Adding additional constraints to the fit requiring $\int p(x|\theta)\ dx$ to be 1 at $\theta=0$ and $\theta=1$ yields the orange dashed line PDF instead, which is a good approximation of the true distribution.}
    \label{fig:fit_bias}
\end{figure}

To understand the source of the problem, it is helpful to consider the joint probability density, which is visualized in Fig.~\ref{fig:fit_bias}. We know the ``true'' value we want to reproduce, which is the product of the distribution $p(x|\theta)$ in Eq.~\eqref{eq:distribution} with a flat prior $p(\theta)=1$ on the range $\theta \in [0,1]$. We can then overlay $\hat f(x,\theta)$ given by the best-fit values of $\mu_m$, $\mu_b$, and $\sigma$. We see that the overall shapes of the two are similar, but $\hat f(x,\theta)$ is more concentrated at intermediate values of $\theta$. Integrating along vertical lines, the value $\int \hat f(x|\theta=0)\ dx$ is clearly less than $\int \hat f(x|\theta=0.5)\ dx$, while the values $\int p(x|\theta)\ dx$ are constant for all $\theta$. While maintaining the \textit{total probability} in the joint PDF constant, the fit to $f$ has recognized that a better value of the likelihood can be achieved by removing probability from the regions near $\theta=0$ and $\theta=1$ (where there are no data points to enter the likelihood computation) and placing that likelihood near $\theta=0.5$, a parameter value that is consistent with almost all values of $x$. In other words, the function $\hat f$ is not consistent with the presumption of a flat prior $p(\theta)$.

Having seen this, we can now propose a solution: we require the fit to minimize the cost in Eq.~\eqref{eq:likelihood} while also satisfying the following constraint:
\begin{equation}
    \int_{-\infty}^{\infty} p \left( \theta_j, x \right) dx = 1 \, \forall \, \theta_j \in \left[ 0, 1 \right].
\end{equation}
This constraint prevents the optimizer from improving the cost function by implicitly modifying $p(\theta)$. In practice, the constraint is applied for a few values of $\theta$, for example $\theta=0, 0.5, 1$, as this was seen to be enough to correct the bias. In this case, the smoothness of the functional forms forced the constraint to apply everywhere. Applying this constraint, we redo the fit. The best-fit parameters obtained---the bottom row of Table \ref{tab:fitresults}---are now consistent with the true values used to generate the data. Accordingly, the joint PDF of the constrained fit (in Fig. \ref{fig:fit_bias}) shows no bias towards $\theta=0.5$. Any remaining disagreement is attributable to statistical fluctuations in the estimation of $p(x)$ from the finite data set.

\subsection{MDN With One Observable}\label{s:example1d}
Now we will demonstrate how this correction is applied with a Mixture Density Network. We construct a neural network using PyTorch~\cite{NEURIPS2019_9015} with a single input $x$, and outputs which are parameters of a function $\tilde p \left( \theta | x \right)$ which estimates the likelihood $\mathcal{L} \left( \theta | x \right)$. (See \S\ref{sec:mdn} for details.) There is a single hidden layer with an adjustable number of nodes. Because of the intentional simplicity of these data, the number of nodes in this hidden layer was generally kept between 2 and 5, and we model the posterior density with just a single Gaussian function. Therefore, the two output nodes of the network are just the mean and standard deviation of this function.\footnote{To avoid ambiguity, it is common practice in MDN applications that the output node corresponding to the Gaussian's $\sigma$ is actually $\log \left( \sigma \right)$, and the exponential of this node's value is used to calculate the loss. This action effectively forces $\sigma>0$.}

\begin{figure}
    \centering
    \includegraphics[width=0.4\textwidth]{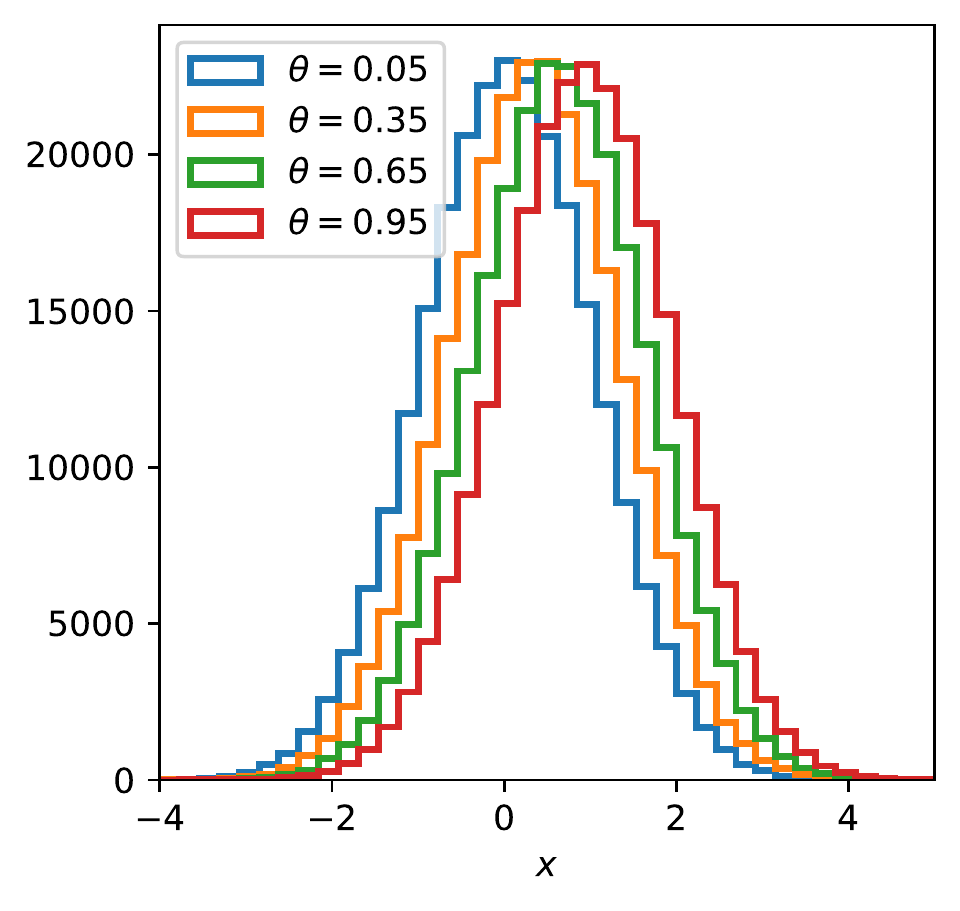}
    \caption{Four of the ten template histograms used as training data for the single-input MDN of \S\ref{s:example1d}.}
    \label{fig:1d_generator_histograms}
\end{figure}

\begin{figure}
    \centering
    \includegraphics[width=0.48\textwidth]{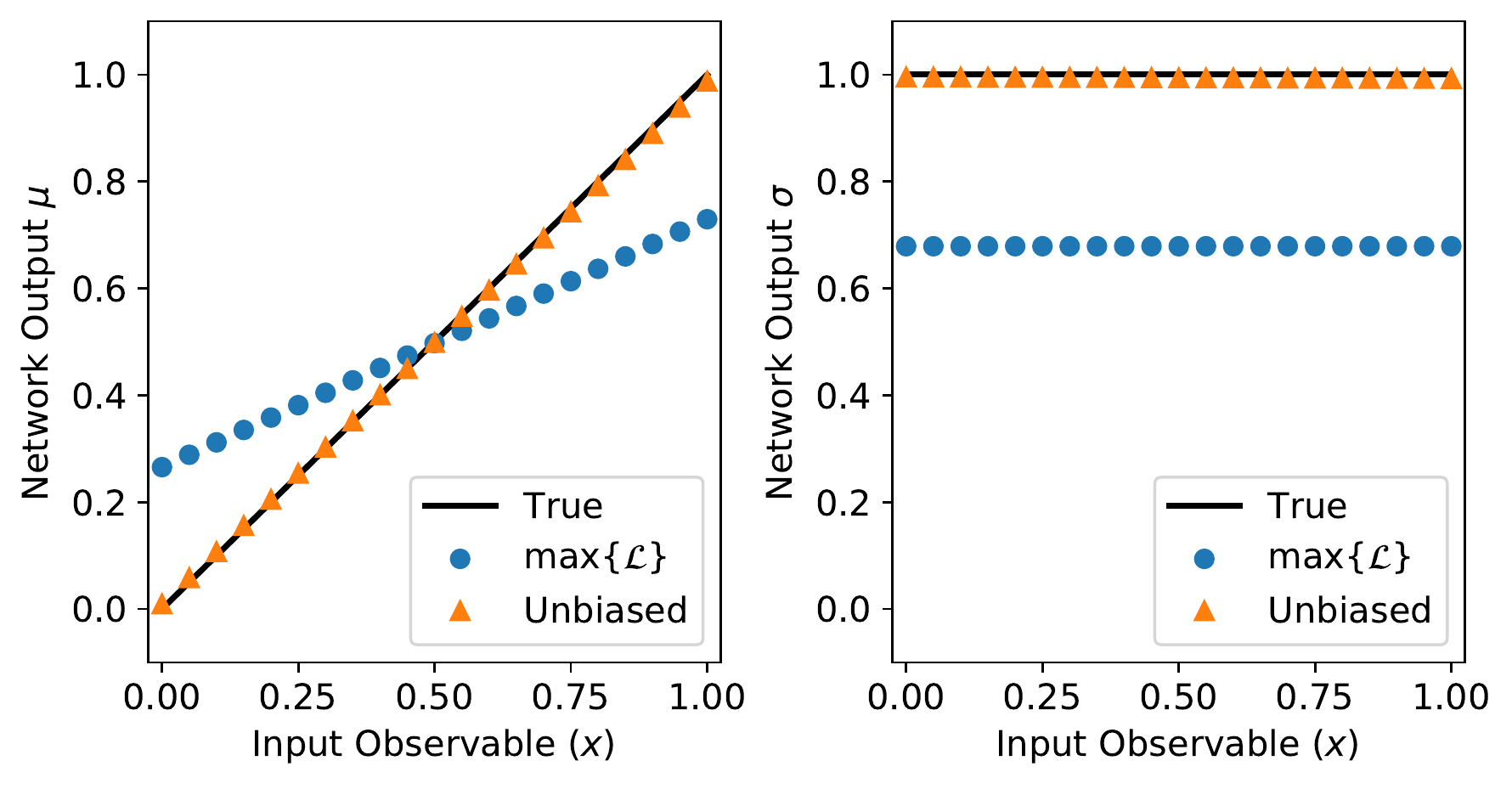}
    \caption{Parameters of the posterior density functions predicted by the MDNs of \S\ref{s:example1d} for various values of the input $x$. Without correction, the network will underestimate the width of the posterior density, and predict a maximum-likelihood $\theta$ value biased towards the middle of the sampled range.}
    \label{fig:net1D_node_scan}
\end{figure}

A large number of $x$ values were generated at each of 10 equally-spaced discrete values of $\theta$ from Gaussian distributions with $\mu=\theta$ and $\sigma=1$. The network is trained by minimizing the cost function described in \S\ref{sec:mdn}. At each epoch, the cost is backpropogated through the network, and the Adam minimizer~\cite{kingma2017adam} is used to adjust the network parameters. We find that this produces biased results. Due to the construction of the training data, we expect that the MDN should predict that $\tilde p \left( \theta | x \right)$ is a truncated Gaussian (nonzero only on the range $\theta \in \left[ 0, 1 \right]$) with $\mu=x$ and $\sigma=1$. In Fig.~\ref{fig:net1D_node_scan}, the MDN output from the naive training for input values of $x$ are shown; we see that the network underestimates the width of the posterior density and biases the mean towards $0.5$.

\subsubsection{1D Network Correction and Validation}
The cost function of the network is then modified by adding an extra term,
\begin{align} \label{eq:unbiased_likelihood}
    \mathcal{C} \rightarrow \mathcal{C} + \lambda \mathcal{S},
\end{align} where
\begin{equation} \label{eq:unbiased_likelihood_std}
    \mathcal{S} = \text{std.} \left\{ \int_{-\infty}^{\infty} \tilde p \left( \theta_j, x \right) dx \right\},
\end{equation}
for some set of parameter values $\theta_j$. With respect to Fig. \ref{fig:fit_bias}, this extra term is the standard deviation of integrals of the joint probability $\tilde p \left(x, \theta \right)$ along vertical slices in $\theta$. This term ensures that these integrals are all the same, thus ensuring that $p(x | \theta)$ remains constant for any value of $\theta$, discouraging a bias towards intermediate values of $\theta$. The joint probability,
\begin{align*}
    \tilde p(\theta, x) = \tilde p(\theta|x) p(x),
\end{align*}
is estimated from the product of the MDN output and the probability distribution over the training sample, which is estimated using histograms.

The new hyperparameter $\lambda$ is tuned to balance the effectiveness of the additional term while avoiding the introduction of numerical instability into the minimization. In practice, its value during the training is initially set to zero, allowing the network to first loosely learn the posterior. Then, in a second step of the training, its value is increased so that the values of the two terms in Eq.~\ref{eq:unbiased_likelihood} are approximately equal. With this new cost function, the outputs of the neural network---the coefficients of the Gaussian posterior density function---correctly match expectations, as seen in Fig.~\ref{fig:net1D_node_scan}.

\begin{figure}
    \centering
    \includegraphics[width=0.4\textwidth]{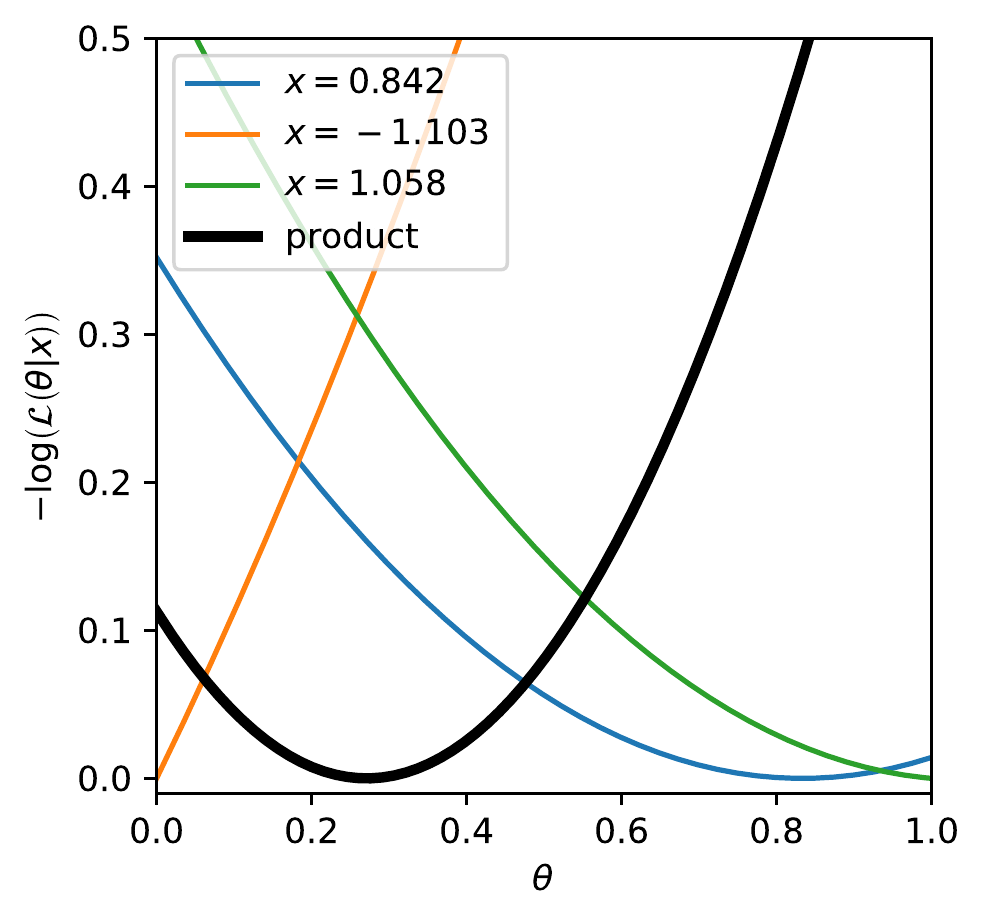}
    \caption{For a fixed value of the parameter $\theta=0.2$, three $x_i$ values are generated. The posteriors $\tilde{p}_i \left( \theta | x_i \right)$ are obtained from the output of the trained MDN of \S\ref{s:example1d}. The maximum likelihood estimator $\theta_\textrm{ML}$ is extracted from the product of these posteriors according to Eq.~\eqref{eq:net1D_theta_ml} and shows a result statistically compatible with the expected value of 0.2. (For visualization, all functions are offset to have a minimum of 0.)}
    \label{fig:net1D_likelihoods}
\end{figure}

Next, we demonstrate the ability of the network to reconstruct the underlying parameter $\theta$ given a set of observations. Each measurement of the quantity $x$ will be denoted $x_i$. For each $x_i$, the trained network outputs an estimate of the posterior function $\tilde{p}_i \left( \theta |x_i \right)$. To calculate the combined cost of an entire set of measurements, $\left\{ x_i \right\}$, we take the negative logarithm of the product of the individual posteriors,
\begin{align}
    \mathcal{C} \left( \theta | \left\{ x_i \right\} \right) = - \log \left[ \prod_{i} \tilde{p}_i \left( \theta |x_i \right) \right].
\end{align}
The maximum likelihood estimator is defined as
\begin{equation} \label{eq:net1D_theta_ml}
    \theta_\textrm{ML} = \argmin_{\theta \in [0, 1]} \left\{ \mathcal{C} \left( \theta | \left\{ x_i \right\} \right) \right\}
\end{equation}
A likelihood-ratio test can be used to determine the statistical uncertainty of the estimator $\theta_\textrm{ML}$. An example of the functions generated for this trained network is given in Fig.~\ref{fig:net1D_likelihoods}.

\begin{figure}
    \centering
    \includegraphics[width=0.4\textwidth]{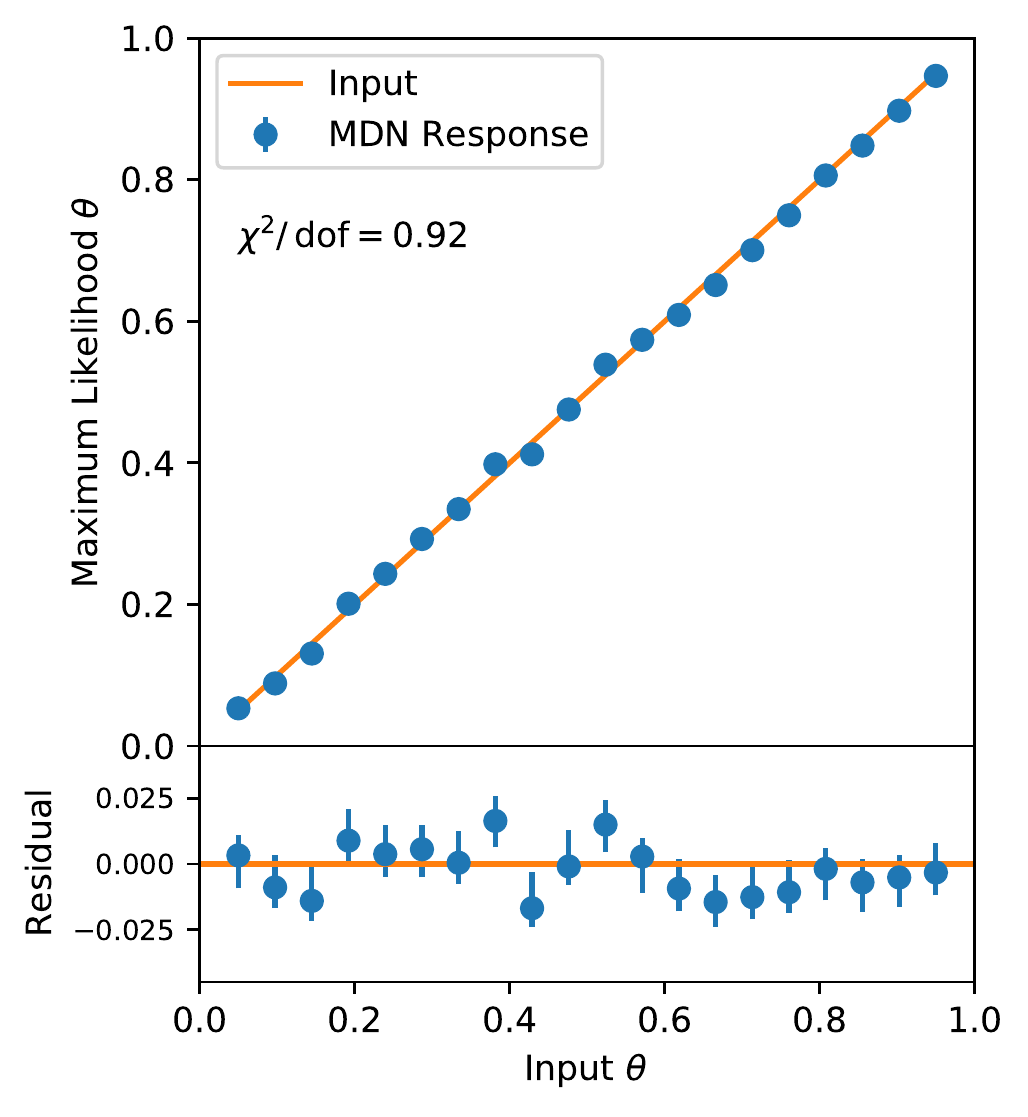}
    \caption{Comparison of the true value of the parameter $\theta$ and the MLE estimate determined for a dataset generated at that $\theta$ over the range $\theta \in [0,1]$. The errors shown are those obtained by applying Wilks' theorem to the posterior density. No bias is observed and the error estimates meet expectations.}
    \label{fig:net1D_responses}
\end{figure}

To test the accuracy and precision of the network, we choose 20 equally-spaced test $\theta_t$ values in the range $[0,1]$. For each $\theta_t$, we generate a set of 10,000 measurements $\left\{ x_i \right\}$. The pseudodata is passed though the network for each $\theta_t$, the posterior density functions calculated, and a $\theta_\textrm{ML}$ is found. The uncertainty in $\theta_\textrm{ML}$ is estimated using Wilks' theorem \cite{Wilks_wilks-theorem} (searching for values of $\theta$ that increase $\mathcal{C}$ by 0.5). If the network is accurate and unbiased, we should find statistical agreement between $\theta_t$ and $\theta_\textrm{ML}$. Indeed, we found this to be true with $\chi^2/\textrm{d.o.f.} = 0.92$. The results for the unbiased network are shown in Fig. \ref{fig:net1D_responses}.

\subsection{Demonstration With 2D Network} \label{s:example2d}
Next, we consider a more advanced usage of MDNs. We consider a system with two inputs, $\mathbf{x} = \left( x_1, x_2 \right)$. The value of $x_1$ is sensitive to an underlying parameter $\theta$, but only for small values of $x_2$; when $x_2$ is large, there is no sensitivity. This is chosen to demonstrate how this neural network technique can be an advantageous approach to parameter estimation: the MDN will automatically learn when observables offer meaningful information on parameters and when they do not.

\subsubsection{2D Toy Model}
The model used to generate 2D pseudodata comprises two components. The first is events with a small $x_2$ value. These events have an $x_1$ value which depends on the unknown parameter $\theta$. In the toy model, this component is modeled by a Gaussian in $x_1$. The mean of this Gaussian varies with $\theta$. In $x_2$, it is modeled by a Gaussian with a (lower) mean of 1.

In the other component, the value of $x_2$ is large, and the $x_1$ value is not related to the unknown parameter. This component is modeled with a Gamma distribution in $x_1$, and a Gaussian with a (higher) mean of 3 in $x_2$. The relative fraction of the two components is the final model parameter.

\begin{figure}
    \centering
    \includegraphics[width=0.4\textwidth]{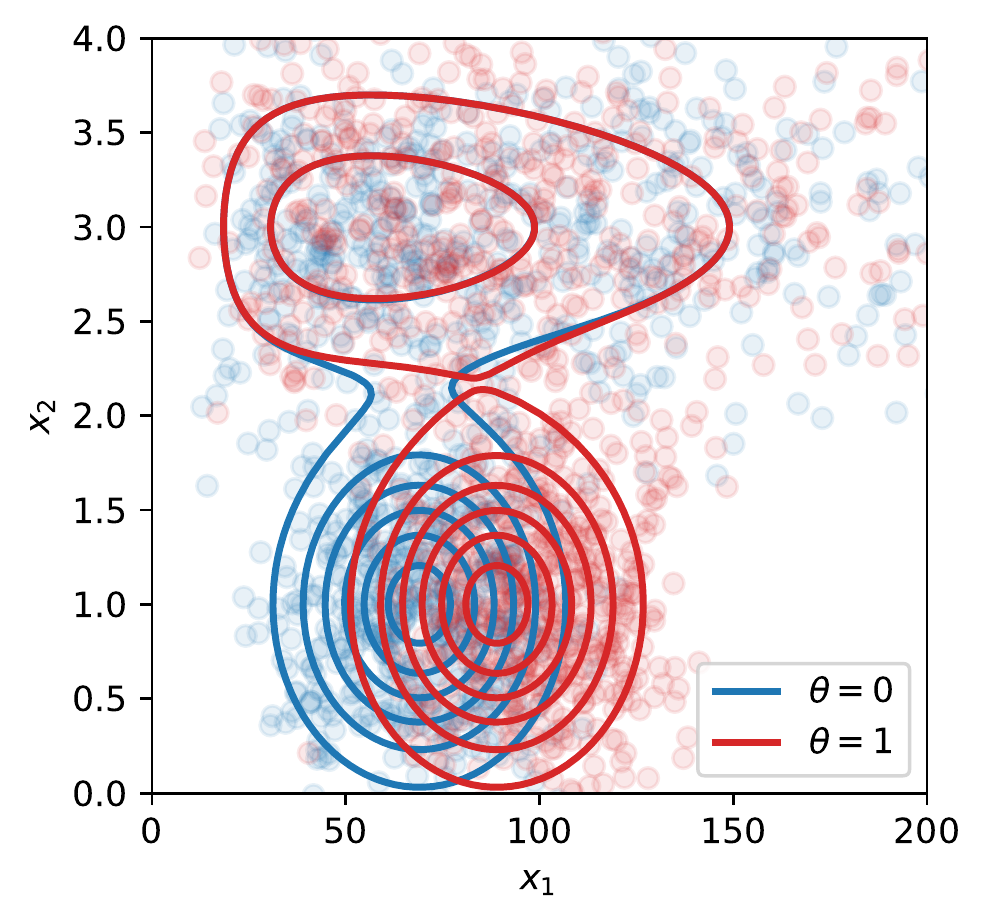}
    \caption{Examples of PDFs (solid lines) and generated sample points (translucent circles) for different values of $\theta$ for the model of \S\ref{s:example2d}.}
    \label{fig:net2D_templates}
\end{figure}

Written out, the 2D PDF takes the form
\begin{align*}
    p\left( x_1, x_2 | \theta \right) = f \cdot p_a\left( x_1, x_2 | \theta \right) + (1-f) \cdot p_b\left( x_1, x_2 \right),
\end{align*} where 
\begin{align*}
    p_a \left( x_1, x_2 | \theta \right) &= \mathcal{N} \left(x_1; \mu_{a1} \left( \theta \right), \sigma_{a1} \right) \cdot \mathcal{N} \left(x_2; \mu_{a2}, \sigma_{a2} \right) \\
    p_b \left( x_1, x_2 \right) &= \Gamma \left(x_1; \mu_{b1}, \alpha, \beta \right) \cdot \mathcal{N} \left(x_2; \mu_{b2}, \sigma_{b2} \right).
\end{align*}
This PDF is shown for the two extremal values of $\theta$ in Fig. \ref{fig:net2D_templates}.

\subsubsection{2D Network Structure, Correction, and Validation}
The network is built in the following manner. First, we note that there are two inputs. Accordingly, there will be two input nodes. Next, one should note from the contours of $p \left( x_1, x_2 \right)$ in Fig. \ref{fig:net2D_templates}, for a sample $\mathbf{x}$ with large $x_2$, the value of $x_1$ has no predictive power of $\theta$. Restated simply, the red and blue contours overlap entirely on the upper half of the plot. However, for small values of $x_2$, the $x_1$ variable can distinguish different values of $\theta$; the red and blue contours are separated on the lower half of the plot. Two hidden layers are used. The output of the network is the coefficients of a mixture of Gaussian distributions. The number of Gaussians in the mixture is a hyperparameter which may be tuned to best match the particular structure of some data. In this example, the data are bimodal, so it is reasonable to expect two Gaussians to be expected. It was confirmed by trial-and-error that a mixture of two Gaussians was adequate to model these data, since it produced the correct linear response in the validation of the network. Therefore, for this network, there are six output nodes (with one normalization constraint) for a total of five independent values. The correction term presented in Eq.~\eqref{eq:unbiased_likelihood} is applied as well. To do this, $p(\bx)$ is estimated using a 2D histogram. It is then multiplied by the MDN output posterior $\tilde p \left( \theta | \bx \right)$ to obtain the joint probability $\tilde p \left( \theta, \bx \right)$. Finally, the integrals in Eq.~\eqref{eq:unbiased_likelihood_std} are calculated. The training data consist of 100,000 samples generated according to $p \left( x_1, x_2 | \theta \right)$ at each of 10 equally-spaced discrete values of $\theta$. 

\begin{figure}
    \centering
    \includegraphics[width=0.4\textwidth]{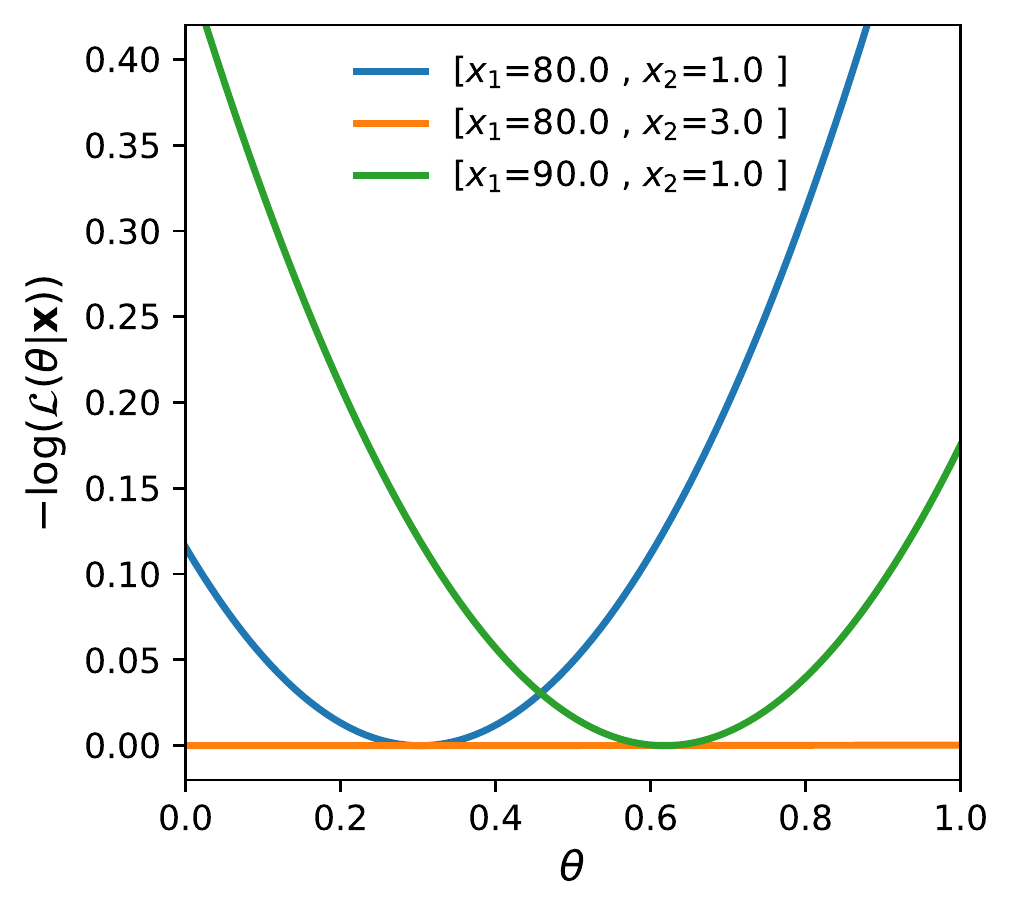}
    \caption{Posterior densities predicted by the trained MDN for three possible observations in the model of \S\ref{s:example2d}. When $x_2$ is small (blue and green curves), the value of $x_1$ gives sensitivity to the model parameter $\theta$. When $x_2$ is large (orange curve), there is very little sensitivity to $\theta$ and the posterior is flat. (For visualization, all functions are offset to have a minimum of 0.)}
    \label{fig:net2D_likelihoods}
\end{figure}

Fig.~\ref{fig:net2D_likelihoods} demonstrates that the trained network is able to learn that the sensitivity of individual observations to the value of the model parameter $\theta$ varies with $x_2$; the predicted posterior density is much flatter in regions where all parameter values give similar distributions.

\begin{figure}
    \begin{center}
    \includegraphics[width=0.4\textwidth]{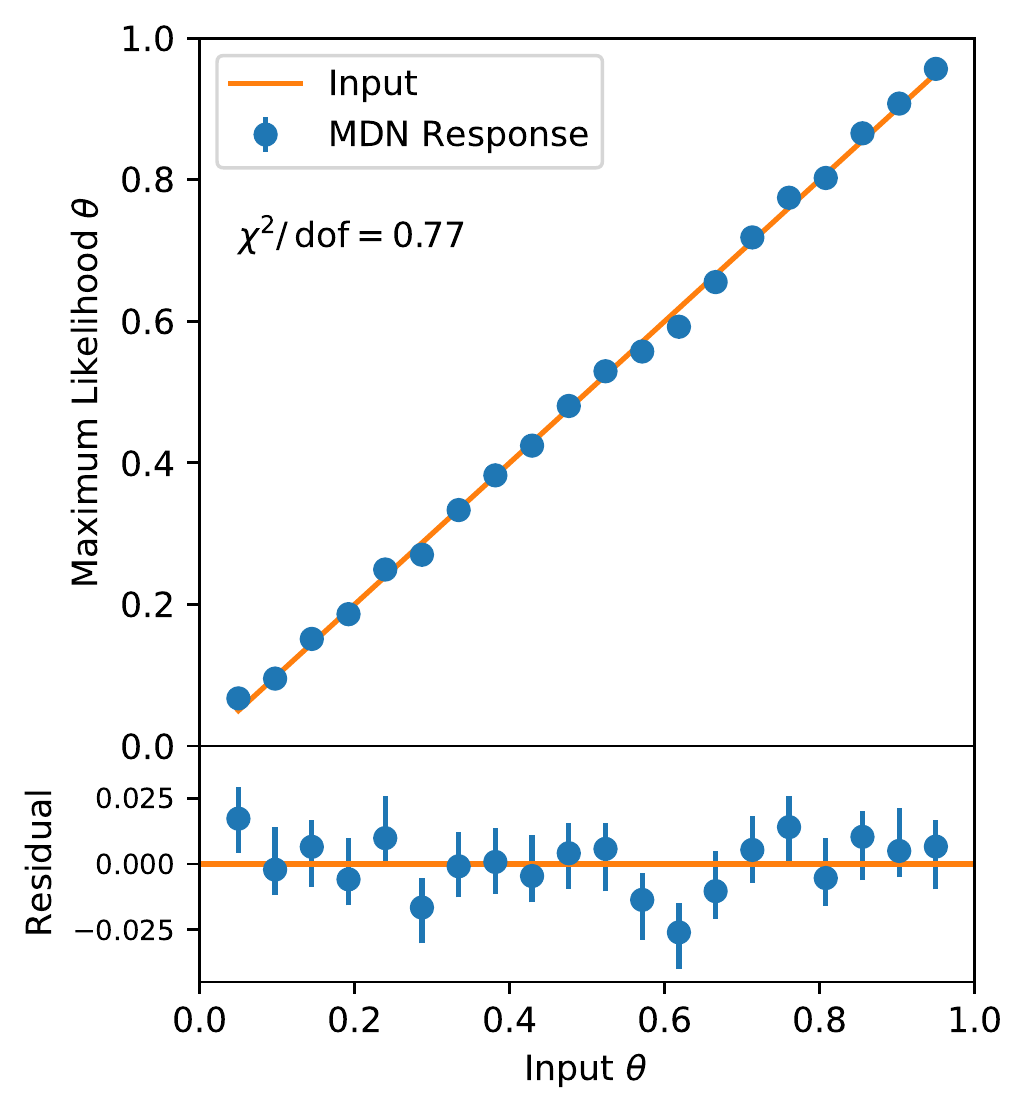}
    \caption{Maximum likelihood estimator for parameter value $\theta_\textrm{ML}$, and associated statistical uncertainties estimated using Wilks' theorem, for samples generated at various values of true $\theta$. No bias is observed and the error estimation works well.}
    \label{fig:net2D_responses}
    \end{center}
\end{figure}

To validate the network's performance, we generate test data sets $\left\{ \mathbf{x}_i \right\}$ for various test parameter values $\theta_t$. As in the one observable case of \S\ref{s:example1d}, each point $\mathbf{x}_i$ is passed through the network and a function $\tilde{p}_i \left( \theta | \mathbf{x}_i \right)$ is calculated. We then find the maximum likelihood estimator of $\theta$. An accurate network should, within statistical uncertainty, reproduce the line $\theta_\mathrm{ML} = \theta_t$. Indeed, Fig. \ref{fig:net2D_responses} shows that the network is accurate and unbiased, and that it provides a reasonable uncertainty estimate.

\section{Method Limitations}
In the previous section, we have demonstrated necessary corrections for networks taking one or two values for each observation. In principle, this method could work with an arbitrarily high number of input observables. However, the current correction technique has a curse-of-dimensionality problem in that computing the correction term $\mathcal{S}$ defined in Eq.~\eqref{eq:unbiased_likelihood_std} requires an estimate of $p(\bx)$ over the full training data set. When determined using histograms with $m$ bins per axis, $N$ observables will require an $m^N$-bin histogram to be reasonably well populated. Other techniques such as generic kernel density estimation~\cite{parzen1962} or $k$-nearest neighbors run into the same issues eventually, although at differing rates, and the best option should be explored in each specific situation. The dimensionality issue is alleviated somewhat by the fact that only one function ($p(\bx)$ over the entire training dataset) needs to be estimated, unlike methods which need to generate templates separately at each generated parameter point.

Another limitation concerns the spacing of training points. While we have demonstrated that MDNs can be trained with discrete templates of data and still interpolate properly to the full continuous parameter space, it has been necessary to use templates from parameters which are uniformly distributed, to enforce that $p(\btheta)$ is flat. If the parameter values are not equally spaced, this would correspond to a non-flat $p(\btheta)$. In principle, weights could be used during training to correct for this effect, but reconstructing $p(\btheta)$ from the distribution of \btheta is a density estimation problem and it is not clear if there is an optimal estimation method for the required $p(\btheta)$ if there are only a limited number of \btheta available. This is an area for future investigation.

\section{Conclusions}
Mixture Density Networks can output posterior density functions which robustly approximate likelihood functions and enable the estimation of parameters from data, even if the training data is only available at discrete values. This method permits one to proceed directly from (possibly multiple) observables to a likelihood function without having to perform the intermediate step of creating a statistical model for the observables, as would be required by many parameter estimation techniques. The MDN technique can be applied even with training data that provide a discrete set of parameter points, provided that the points are spaced evenly and certain corrections and constraints are applied during the training of the network.

An introductory tutorial has been implemented in a Jupyter Notebook and made public~\cite{mdn_tutorial}.

\begin{acknowledgements}
This work was supported by the US Department of Energy, Office of Science, Office of High Energy Physics, under Award Number DE-SC0007890.
\end{acknowledgements}

\bibliographystyle{spphys}
\bibliography{bibliography}

\end{document}